\documentclass[prd,onecolumn,nofootinbib]{revtex4}
\usepackage{bm,amsmath,amssymb,graphicx}

\begin{document}

\noindent
Classical and Quantum Gravity {\bf 31} (6), 065005 (2014)\\

\title{The energy and momentum of the Universe}
\author{Nikodem J. Pop{\l}awski}

\altaffiliation{NPoplawski@newhaven.edu}
\affiliation{Department of Mathematics and Physics, University of New Haven, West Haven, CT, USA}

\begin{abstract}
The Einstein--Cartan--Sciama--Kibble theory of gravity naturally extends general relativity to include quantum-mechanical, intrinsic angular momentum of matter by equipping spacetime with torsion.
Using the Einstein energy--momentum pseudotensor for the gravitational field in this theory, we show that the energy and momentum of the closed Universe are equal to zero.
Since the positive energy from mass and motion of the observed matter in the Universe exceeds in magnitude the negative energy from gravity, the Universe must contain another form of matter whose energy is negative.
This form, which cannot be composed from particles, might be the observed dark matter.\\ \\
Keywords: torsion, spin, Universe, dark matter, black hole.
\end{abstract}
\maketitle

\noindent
{\bf 1. Gravity with torsion}\\ \\
The intrinsic angular momentum (spin) of matter in curved spacetime requires the metric-torsion formulation of gravity, in which the anti-symmetric part of the affine connection (the torsion tensor) is not constrained to be zero but is a variable in the principle of stationary action \cite{tor}.
Regarding the metric and torsion tensors (or the tetrad and spin connection) as independent variables gives the correct generalization of the conservation law for the total (orbital plus intrinsic) angular momentum to the presence of the gravitational field.
The metric-torsion formulation extends general relativity to the simplest theory of gravity with intrinsic spin: the Einstein--Cartan--Sciama--Kibble (ECSK) theory \cite{SK,Lord}.
The coupling between fermions and torsion in this theory avoids the big-bang singularity, replacing it with a bounce at a finite minimum scale factor, and explains why the observable Universe at largest scales appears spatially flat, homogeneous, and isotropic \cite{cosm}.
It may also avoid singularities in black holes \cite{infl}.
In addition, fermions in spacetime with torsion are spatially extended, which may provide a natural ultraviolet cutoff for their propagators in quantum field theory \cite{non}.

The action for the gravitational field and matter in the ECSK theory is equal to
\begin{equation}
S=\frac{1}{c}\int\biggl(-\frac{1}{2\kappa}R\sqrt{-\mathfrak{g}}+\mathfrak{L}_\textrm{m}\biggr)d\Omega,
\label{action1}
\end{equation}
where $R=R_{ik}g^{ik}$ is the curvature scalar, $R_{ik}=R^j_{\phantom{j}ijk}$ is the Ricci tensor, $R^i_{\phantom{i}mjk}=\partial_{j}\Gamma^{\,\,i}_{m\,k}-\partial_{k}\Gamma^{\,\,i}_{m\,j}+\Gamma^{\,\,i}_{l\,j}\Gamma^{\,\,l}_{m\,k}-\Gamma^{\,\,i}_{l\,k}\Gamma^{\,\,l}_{m\,j}$ is the curvature tensor, $\mathfrak{g}$ is the determinant of the metric tensor $g_{ik}$, $\mathfrak{L}_\textrm{m}$ is the Lagrangian density for matter, and $d\Omega$ is the element of four-volume.
We use the notations of \cite{tor,Niko}.
The metricity condition $g_{ij;k}=0$, where semicolon denotes the covariant derivative with respect to the affine connection, gives $\Gamma^{\,\,k}_{i\,j}=\{^{\,\,k}_{i\,j}\}+C^k_{\phantom{k}ij}$, where $\{^{\,\,k}_{i\,j}\}=(1/2)g^{km}(g_{mi,j}+g_{mj,i}-g_{ij,m})$ are the Christoffel symbols, $C^i_{\phantom{i}jk}=S^i_{\phantom{i}jk}+S_{jk}^{\phantom{jk}i}+S_{kj}^{\phantom{kj}i}$ is the contortion tensor, and $S^i_{\phantom{i}jk}=\Gamma^{\,\,\,\,i}_{[j\,k]}$ is the torsion tensor.
The curvature tensor can be decomposed as $R^i_{\phantom{i}klm}=P^i_{\phantom{i}klm}+C^i_{\phantom{i}km:l}-C^i_{\phantom{i}kl:m}+C^j_{\phantom{j}km}C^i_{\phantom{i}jl}-C^j_{\phantom{j}kl}C^i_{\phantom{i}jm}$, where $P^i_{\phantom{i}mjk}=\partial_{j}\{^{\,\,i}_{m\,k}\}-\partial_{k}\{^{\,\,i}_{m\,j}\}+\{^{\,\,i}_{l\,j}\}\{^{\,\,l}_{m\,k}\}-\{^{\,\,i}_{l\,k}\}\{^{\,\,l}_{m\,j}\}$ is the Riemann tensor and colon denotes the covariant derivative with respect to the Levi-Civita connection (the Christoffel symbols).
Using the Gau\ss--Stokes theorem, $\int V^i_{\phantom{i}:i}\sqrt{-\mathfrak{g}}d\Omega=\oint V^i \sqrt{-\mathfrak{g}}dS_i$, where $dS_i$ is the element of the closed hypersurface surrounding the integration four-volume \cite{LL2}, the action (\ref{action1}) can be written as
\begin{eqnarray}
& & S=\frac{1}{c}\int\biggl(-\frac{1}{2\kappa}\Bigl(P-g^{ik}(2C^l_{\phantom{l}il:k}+C^j_{\phantom{j}ij}C^l_{\phantom{l}kl}-C^l_{\phantom{l}im}C^m_{\phantom{m}kl})\Bigr)\sqrt{-\mathfrak{g}}+\mathfrak{L}_\textrm{m}\biggr)d\Omega \nonumber \\
& & =\frac{1}{c}\int\biggl(-\frac{1}{2\kappa}\Bigl(P-g^{ik}(C^j_{\phantom{j}ij}C^l_{\phantom{l}kl}-C^l_{\phantom{l}im}C^m_{\phantom{m}kl})\Bigr)\sqrt{-\mathfrak{g}}+\mathfrak{L}_\textrm{m}\biggr)d\Omega+\frac{1}{\kappa c}\oint C^{lk}_{\phantom{lk}l}\sqrt{-\mathfrak{g}}dS_k,
\label{action2}
\end{eqnarray}
where $P=P_{ik}g^{ik}=P^j_{\phantom{j}ijk}g^{ik}$ is the Riemannian curvature scalar.
The stationarity of action $\delta S=0$ with respect to the variations of the variables, together with a condition that these variations vanish at the boundary of the integration four-volume, gives the field equations.
Accordingly, the hypersurface integral in (\ref{action2}) does not contribute to the field equations and can be omitted.

The stationarity of action (\ref{action2}) with respect to the variation of the torsion tensor is equivalent to
\begin{equation}
\frac{\delta}{\delta S^j_{\phantom{j}kl}}\biggl(-\frac{1}{2\kappa}\Bigl(P-g^{np}(C^i_{\phantom{i}ni}C^m_{\phantom{m}pm}-C^i_{\phantom{i}nm}C^m_{\phantom{m}pi})\Bigr)\sqrt{-\mathfrak{g}}+\mathfrak{L}_\textrm{m}\biggr)=0.
\end{equation}
This relation gives the Cartan equations \cite{SK,Lord,Niko}:
\begin{equation}
S^j_{\phantom{j}ik}-S_i \delta^j_k+S_k \delta^j_i=-\frac{\kappa}{2}s^{\phantom{ik}j}_{ik},
\label{Cartan}
\end{equation}
where $S_i=S^k_{\phantom{k}ik}$ and $s_{ij}^{\phantom{ij}k}=(2/\sqrt{-\mathfrak{g}})(\delta\mathfrak{L}_\textrm{m}/\delta C^{ij}_{\phantom{ij}k})$ is the spin tensor of matter.
The stationarity of action (\ref{action2}) with respect to the variation of the metric tensor is equivalent to
\begin{equation}
\frac{\delta}{\delta g^{jl}}\biggl(-\frac{1}{2\kappa}\Bigl(P-g^{np}(C^i_{\phantom{i}ni}C^m_{\phantom{m}pm}-C^i_{\phantom{i}nm}C^m_{\phantom{m}pi})\Bigr)\sqrt{-\mathfrak{g}}+\mathfrak{L}_\textrm{m}\biggr)=0.
\label{station1}
\end{equation}
This relation gives the Einstein equations:
\begin{equation}
G_{ik}=\kappa(T_{ik}+U_{ik}),
\label{Einstein}
\end{equation}
where $G_{ik}=P_{ik}-(1/2)Pg_{ik}$ is the Einstein tensor, $T_{ik}=(2/\sqrt{-\mathfrak{g}})(\delta\mathfrak{L}_\textrm{m}/\delta g^{ik})$ is the symmetric, metric energy--momentum tensor of matter, and
\begin{equation}
U_{ik}=\frac{1}{\kappa}\biggl(C^j_{\phantom{j}ij}C^l_{\phantom{l}kl}-C^l_{\phantom{l}ij}C^j_{\phantom{j}kl}-\frac{1}{2}g_{ik}(C^{jm}_{\phantom{jm}j}C^l_{\phantom{l}ml}-C^{mjl}C_{ljm})\biggr).
\label{correct}
\end{equation}
Using the Cartan equations (\ref{Cartan}), the tensor $U^{ik}$ can be written as
\begin{equation}
U^{ik}=\kappa\biggl(-s^{ij}_{\phantom{ij}[l}s^{kl}_{\phantom{kl}j]}-\frac{1}{2}s^{ijl}s^k_{\phantom{k}jl}+\frac{1}{4}s^{jli}s_{jl}^{\phantom{jl}k}+\frac{1}{8}g^{ik}(-4s^l_{\phantom{l}j[m}s^{jm}_{\phantom{jm}l]}+s^{jlm}s_{jlm})\biggr).
\end{equation}
The contracted Bianchi identities for the Einstein tensor, $G^{ik}_{\phantom{ik}:k}=0$, give the Riemannian conservation law for the combined energy--momentum density:
\begin{equation}
\bigl(\sqrt{-\mathfrak{g}}(T^{ik}+U^{ik})\bigr)_{:k}=0.
\label{conserv}
\end{equation}

The scalar density $\sqrt{-\mathfrak{g}}P$ can be written as
\begin{equation}
\sqrt{-\mathfrak{g}}P=\mathcal{G}+(\mathfrak{g}^{ik}\{^{\,\,l}_{i\,k}\})_{,l}-(\mathfrak{g}^{ik}\{^{\,\,l}_{i\,l}\})_{,k},
\end{equation}
where
\begin{equation}
\mathcal{G}=\mathfrak{g}^{ik}(\{^{\,\,m}_{i\,l}\}\{^{\,\,l}_{m\,k}\}-\{^{\,\,m}_{i\,k}\}\{^{\,\,l}_{m\,l}\})
\label{noncov}
\end{equation}
and $\mathfrak{g}^{ik}=\sqrt{-\mathfrak{g}}g^{ik}$.
Using the Gau\ss--Stokes theorem, $d\Omega(\partial/\partial x^i)\leftrightarrow dS_i$, the action for the gravitational field and matter (\ref{action3}) becomes
\begin{equation}
S=\frac{1}{c}\int\biggl(-\frac{1}{2\kappa}\Bigl(\mathcal{G}-{\sf g}^{np}(C^i_{\phantom{i}ni}C^m_{\phantom{m}pm}-C^i_{\phantom{i}nm}C^m_{\phantom{m}pi})\Bigr)+\mathfrak{L}_\textrm{m}\biggr)d\Omega-\frac{1}{2\kappa c}\oint\mathfrak{g}^{ik}\{^{\,\,l}_{i\,k}\}dS_l+\frac{1}{2\kappa c}\oint\mathfrak{g}^{ik}\{^{\,\,l}_{i\,l}\}dS_k.
\label{action3}
\end{equation}
The hypersurface integrals in (\ref{action3}) do not contribute to the field equations and can be omitted.
Accordingly, the stationarity condition (\ref{station1}) is equivalent to
\begin{equation}
\frac{\delta}{\delta g^{jl}}\biggl(-\frac{1}{2\kappa}\Bigl(\mathcal{G}-{\sf g}^{np}(C^i_{\phantom{i}ni}C^m_{\phantom{m}pm}-C^i_{\phantom{i}nm}C^m_{\phantom{m}pi})\Bigr)+\mathfrak{L}_\textrm{m}\biggr)=0,
\label{station2}
\end{equation}
which gives the Einstein equations (\ref{Einstein}).\\ \\

\noindent
{\bf 2. The energy--momentum pseudotensor}\\ \\
Because the quantity (\ref{noncov}) depends on the metric tensor $g^{ij}$ and its first derivatives $g^{ij}_{\phantom{ij},k}$, we can construct a canonical energy--momentum density corresponding to the gravitational Lagrangian density $\mathfrak{L}_\textrm{g}=-1/(2\kappa)\mathcal{G}$:
\begin{equation}
\mathfrak{t}^{\phantom{k}i}_k=-\frac{1}{2\kappa}\biggl(\frac{\partial\mathcal{G}}{\partial g^{jl}_{\phantom{jl},i}}g^{jl}_{\phantom{jl},k}-\delta^i_k\mathcal{G}\biggr).
\label{pseudo1}
\end{equation}
This quantity is not a tensor density because $\mathcal{G}$ is not a scalar density.
Its division by $\sqrt{-\mathfrak{g}}$ defines the Einstein energy--momentum pseudotensor for the gravitational field \cite{Lord,Niko,Schr}:
\begin{equation}
\frac{\mathfrak{t}^{\phantom{k}i}_k}{\sqrt{-\mathfrak{g}}}=-\frac{1}{2\kappa}\biggl(\frac{\partial{\sf G}}{\partial g^{jl}_{\phantom{jl},i}}g^{jl}_{\phantom{jl},k}-\delta^i_k{\sf G}\biggr),
\label{pseudo2}
\end{equation}
where ${\sf G}=\mathcal{G}/\sqrt{-\mathfrak{g}}$.
Differentiating (\ref{pseudo1}) gives
\begin{eqnarray}
& & 2\kappa\mathfrak{t}^{\phantom{k}i}_{k\phantom{i},i}=-\partial_i\frac{\partial\mathcal{G}}{\partial g^{jl}_{\phantom{jl},i}}g^{jl}_{\phantom{jl},k}-\frac{\partial\mathcal{G}}{\partial g^{jl}_{\phantom{jl},i}}g^{jl}_{\phantom{jl},ki}+\mathcal{G}_{,k}=-\partial_i\frac{\partial\mathcal{G}}{\partial g^{jl}_{\phantom{jl},i}}g^{jl}_{\phantom{jl},k}-\frac{\partial\mathcal{G}}{\partial g^{jl}_{\phantom{jl},i}}g^{jl}_{\phantom{jl},ki}+\frac{\partial\mathcal{G}}{\partial g^{jl}}g^{jl}_{\phantom{jl},k} \nonumber \\
& & +\frac{\partial\mathcal{G}}{\partial g^{jl}_{\phantom{jl},i}}g^{jl}_{\phantom{jl},ik}=\biggl(\frac{\partial\mathcal{G}}{\partial g^{jl}}-\partial_i\frac{\partial\mathcal{G}}{\partial g^{jl}_{\phantom{jl},i}}\biggr)g^{jl}_{\phantom{jl},k}=\frac{\delta\mathcal{G}}{\delta g^{jl}}g^{jl}_{\phantom{jl},k},
\end{eqnarray}
which, using (\ref{correct}) and (\ref{station2}), leads to
\begin{equation}
\mathfrak{t}^{\phantom{k}i}_{k\phantom{i},i}=\frac{\delta\bigl(\mathfrak{L}_\textrm{m}+\frac{1}{2\kappa}\sqrt{-\mathfrak{g}}g^{np}(C^i_{\phantom{i}ni}C^m_{\phantom{m}pm}-C^i_{\phantom{i}nm}C^m_{\phantom{m}pi})\bigr)}{\delta g^{jl}}g^{jl}_{\phantom{jl},k}=\frac{1}{2}\sqrt{-\mathfrak{g}}(T_{jl}+U_{jl})g^{jl}_{\phantom{jl},k}.
\label{Eps6}
\end{equation}

The Riemannian conservation law (\ref{conserv}) gives
\begin{eqnarray}
& & \bigl(\sqrt{-\mathfrak{g}}(T^{\phantom{k}i}_k+U^{\phantom{k}i}_k)\bigr)_{,i}=\{^{\,\,l}_{k\,i}\}\sqrt{-\mathfrak{g}}(T^{\phantom{l}i}_l+U^{\phantom{l}i}_l)=\frac{1}{2}g^{lm}g_{im,k}\sqrt{-\mathfrak{g}}(T^{\phantom{l}i}_l+U^{\phantom{l}i}_l) \nonumber \\
& & =-\frac{1}{2}g^{lm}_{\phantom{lm},k}\sqrt{-\mathfrak{g}}(T_{lm}+U_{lm}).
\label{Eps7}
\end{eqnarray}
As a result of adding (\ref{Eps6}) and (\ref{Eps7}), the ordinary four-divergence of the total energy--momentum density for the gravitational field and matter vanishes:
\begin{equation}
\bigl(\mathfrak{t}^{\phantom{k}i}_k+\sqrt{-\mathfrak{g}}(T^{\phantom{k}i}_k+U^{\phantom{k}i}_k)\bigr)_{,i}=\biggl(\mathfrak{t}^{\phantom{k}i}_k+\frac{\sqrt{-\mathfrak{g}}}{\kappa}G^{\phantom{k}i}_k\biggr)_{,i}=0.
\label{Eps8}
\end{equation}
Integrating (\ref{Eps8}) over the four-volume and using the Gau\ss--Stokes theorem gives
\begin{equation}
\oint\bigl(\mathfrak{t}^{\phantom{k}i}_k+\sqrt{-\mathfrak{g}}(T^{\phantom{k}i}_k+U^{\phantom{k}i}_k)\bigr)dS_i=0.
\end{equation}
If the hypersurface represented by the element $dS_k$ is taken as a hyperplane perpendicular to the $x^0$ axis (volume hypersurface), $dS_k=\delta_k^0 dV$, then the closed hypersurface surrounds the four-volume between two hyperplanes at times $t_1$ and $t_2$:
\begin{equation}
\oint\bigl(\mathfrak{t}^{\phantom{k}0}_k+\sqrt{-\mathfrak{g}}(T^{\phantom{k}0}_k+U^{\phantom{k}0}_k)\bigr)dV\Big|_{t_1}^{t_2}=0.
\label{fmam2}
\end{equation}
The four-momentum of the gravitational field and matter,
\begin{equation}
P_i=\frac{1}{c}\int\bigl(\mathfrak{t}^{\phantom{i}0}_i+\sqrt{-\mathfrak{g}}(T^{\phantom{i}0}_i+U^{\phantom{i}0}_i)\bigr)dV=\frac{1}{c}\int\bigl(\mathfrak{t}^{\phantom{i}k}_i+\sqrt{-\mathfrak{g}}(T^{\phantom{i}k}_i+U^{\phantom{i}k}_i)\bigr)dS_k,
\label{Eps10}
\end{equation}
which is not a vector, is therefore conserved:
\begin{equation}
P_i|_{t_1}=P_i|_{t_2}=P_i=\mbox{constant}.
\end{equation}

Using $\{^{\,\,m}_{i\,j}\}=-(1/2)(g_{kj}g^{mk}_{\phantom{mk},i}+g_{ki}g^{mk}_{\phantom{mk},j}-g^{mk}g_{il}g_{jn}g^{ln}_{\phantom{ln},k})$ and $\{^{\,\,k}_{k\,i}\}=-(1/2)g_{jk}g^{jk}_{\phantom{jk},i}=(\textrm{ln}\sqrt{-\mathfrak{g}})_{,i}$ gives
\begin{eqnarray}
& & \frac{\partial\{^{\,\,m}_{i\,l}\}}{\partial g^{rs}_{\phantom{rs},n}}=-\frac{1}{2}(g_{l(r}\delta^m_{s)}\delta^n_i+g_{i(r}\delta^m_{s)}\delta^n_l-g^{mn}g_{i(r}g_{s)l}), \\
& & \frac{\partial\{^{\,\,l}_{m\,l}\}}{\partial g^{rs}_{\phantom{rs},n}}=-\frac{1}{2}g_{rs}\delta^n_m.
\end{eqnarray}
Consequently, we obtain
\begin{eqnarray}
& & \frac{\partial{\sf G}}{\partial g^{rs}_{\phantom{rs},n}}=2g^{ik}\{^{\,\,m}_{i\,l}\}\frac{\partial\{^{\,\,l}_{m\,k}\}}{\partial g^{rs}_{\phantom{rs},n}}-g^{ik}\{^{\,\,l}_{m\,l}\}\frac{\partial\{^{\,\,m}_{i\,k}\}}{\partial g^{rs}_{\phantom{rs},n}}-g^{ik}\{^{\,\,m}_{i\,k}\}\frac{\partial\{^{\,\,l}_{m\,l}\}}{\partial g^{rs}_{\phantom{rs},n}} \nonumber \\
& & =-\{^{\,\,n}_{r\,s}\}+\frac{1}{2}\Bigl(\{^{\,\,l}_{s\,l}\}\delta^n_r+\{^{\,\,l}_{r\,l}\}\delta^n_s-\{^{\,\,l}_{m\,l}\}g^{mn}g_{rs}\Bigr)+\frac{1}{2}\{^{\,\,n}_{j\,l}\}g^{jl}g_{rs},
\end{eqnarray}
which leads to
\begin{equation}
\frac{\partial\mathcal{G}}{\partial g^{rs}_{\phantom{rs},i}}g^{rs}_{\phantom{rs},k}=-\sqrt{-\mathfrak{g}}\{^{\,\,i}_{r\,s}\}g^{rs}_{\phantom{rs},k}+\sqrt{-\mathfrak{g}}\{^{\,\,l}_{r\,l}\}g^{ri}_{\phantom{ri},k}+\{^{\,\,l}_{m\,l}\}g^{mi}(\sqrt{-\mathfrak{g}})_{,k}-\{^{\,\,i}_{j\,l}\}g^{jl}(\sqrt{-\mathfrak{g}})_{,k}.
\end{equation}
The Einstein pseudotensor (\ref{pseudo2}) can thus be written as \cite{LL2}
\begin{equation}
\frac{\mathfrak{t}^{\phantom{k}i}_k}{\sqrt{-\mathfrak{g}}}=\frac{1}{2\kappa\sqrt{-\mathfrak{g}}}(\{^{\,\,i}_{l\,m}\}{\sf g}^{lm}_{\phantom{lm},k}-\{^{\,\,l}_{m\,l}\}{\sf g}^{mi}_{\phantom{mi},k}+\delta^i_k\mathcal{G}).
\label{Eps15}
\end{equation}
Accordingly, $\mathfrak{t}_{ik}$ is not symmetric in the indices $i,k$.
The quantity
\begin{equation}
\mathfrak{t}^{\phantom{i}k}_i+\sqrt{-\mathfrak{g}}(T^{\phantom{i}k}_i+U^{\phantom{i}k}_i)=\mathfrak{t}^{\phantom{i}k}_i+\frac{\sqrt{-\mathfrak{g}}}{\kappa}G^{\phantom{i}k}_i
\label{Eps11}
\end{equation}
is the Einstein--Cartan energy--momentum complex \cite{Niko,Gar}.

The conservation law (\ref{Eps8}) infers that the complex (\ref{Eps11}) can be written as
\begin{equation}
\mathfrak{t}^{\phantom{k}l}_k+\frac{\sqrt{-\mathfrak{g}}}{\kappa}G^{\phantom{k}l}_k=\eta^{\phantom{k}li}_{k\phantom{li},i},
\label{Eps17}
\end{equation}
where $\eta^{\phantom{k}li}_k$ satisfies
\begin{equation}
\eta^{\phantom{k}li}_k=-\eta^{\phantom{k}il}_k.
\end{equation}
We define \cite{LL2}
\begin{eqnarray}
& & \lambda^{iklm}=\frac{1}{2\kappa}(-\mathfrak{g})(g^{ik}g^{lm}-g^{il}g^{km}), \label{Eps21} \\
& & h^{ikl}=\lambda^{iklm}_{\phantom{iklm},m}=-h^{ilk}.
\label{Eps22}
\end{eqnarray}
Using $g^{ik}_{\phantom{ik}:j}=g^{ik}_{\phantom{ik},j}+\{^{\,\,i}_{l\,j}\}g^{lk}+\{^{\,\,k}_{l\,j}\}g^{il}=0$, the quantity (\ref{Eps22}) is equal to
\begin{equation}
h^{ikl}=\frac{1}{2\kappa}(-\mathfrak{g})(\{^{\,\,l}_{j\,m}\}g^{km}g^{ij}+\{^{\,\,k}_{m\,n}\}g^{mn}g^{il}-\{^{\,\,m}_{m\,n}\}g^{kn}g^{il}-\{^{\,\,k}_{j\,m}\}g^{lm}g^{ij}-\{^{\,\,l}_{m\,n}\}g^{mn}g^{ik}+\{^{\,\,m}_{m\,n}\}g^{ln}g^{ik}).
\end{equation}
Equations (\ref{noncov}) and (\ref{Eps15}) infer that
\begin{equation}
\eta^{\phantom{i}kl}_i=\frac{1}{\sqrt{-\mathfrak{g}}}h^{\phantom{i}kl}_i
\label{Eps24}
\end{equation}
satisfies (\ref{Eps17}) \cite{Mol}.\\ \\ \\ \\

\noindent
{\bf 3. The four-momentum of the Universe}\\ \\
Substituting (\ref{Eps17}) into (\ref{Eps10}), and using (\ref{Eps11}) and the Stokes theorem, $dS_i(\partial/\partial x^k)-dS_k(\partial/\partial x^i)\leftrightarrow df^\star_{ik}$, where $df^\star_{ik}$ is the element of the closed surface which bounds the hypersurface \cite{LL2}, gives
\begin{equation}
P_i=\frac{1}{c}\int\eta^{\phantom{i}kl}_{i\phantom{kl},l}dS_k=\frac{1}{2c}\int(\eta^{\phantom{i}kl}_{i\phantom{kl},l}dS_k-\eta^{\phantom{i}kl}_{i\phantom{kl},k}dS_l)=\frac{1}{2c}\oint\eta^{\phantom{i}kl}_{i\phantom{kl}}df^\ast_{kl}.
\label{moment1}
\end{equation}
If the hypersurface is a volume hypersurface, then the four-momentum of the gravitational field and matter (\ref{moment1}) can be written as a surface integral:
\begin{equation}
P_i=\frac{1}{2c}\oint\eta^{\phantom{i}0\alpha}_{i\phantom{0\alpha}}df^\ast_{0\alpha}=\frac{1}{c}\oint\eta^{\phantom{i}0\alpha}_{i\phantom{0\alpha}}df_\alpha,
\label{moment2}
\end{equation}
where $df_\alpha$ is the element of the closed surface which bounds the integration volume \cite{LL2} and $\alpha$ denotes the spatial tensor indices.
The same formula for $P_i$, with $\eta^{\phantom{i}kl}_i$ given by (\ref{Eps24}), can be derived using the Landau--Lifshitz energy--momentum pseudotensor \cite{LL2}.
In the spherical coordinates, we have $df_\alpha=n_\alpha r^2 do$, where $r$ is the radial coordinate measured from the origin, $n_\alpha$ is the normal vector to the surface, and $do$ is the element of the solid angle \cite{LL2}.
For the Schwarzschild metric of a mass $m$ at the origin, written in the isotropic spherical coordinates, the four-momentum of the gravitational field and matter (\ref{moment2}) is given by \cite{Lord,LL2}
\begin{equation}
P^0=mc,\,\,\,P^\alpha=0.
\label{energy}
\end{equation}

If we neglect spin, then $U_{ik}=0$ and the ECSK theory reduces to general relativity.
In this case, the Einstein--Cartan energy--momentum complex reduces to the Einstein energy--momentum complex, $\mathfrak{t}^{\phantom{i}k}_i+\sqrt{-\mathfrak{g}}T^{\phantom{i}k}_i$, which remains equal to the right-hand side of (\ref{Eps11}).
The conservation law (\ref{Eps8}) reduces to $(\mathfrak{t}^{\phantom{i}k}_i+\sqrt{-\mathfrak{g}}T^{\phantom{i}k}_i)_{,k}=0$.
The Einstein energy--momentum complex can also be written as (\ref{Eps17}) and the four-momentum of the gravitational field and matter is also given by (\ref{moment2}).

Torsion in the ECSK gravity provides a theoretical explanation for a scenario, according to which every black hole creates inside its event horizon a new, baby universe and becomes an Einstein--Rosen bridge (wormhole) that connects this universe to the parent universe in which the black hole exists \cite{infl,BH}.
At extremely high densities, much larger than nuclear densities, torsion manifests itself as a force that counters gravitational attraction, preventing matter in a black hole from compressing to a singularity \cite{cosm,infl}.
Instead, matter reaches a state of finite, extremely high density, stops collapsing, undergoes a bounce, and starts rapidly expanding as a new universe.
Extremely strong gravitational fields near the bounce cause an intense particle production, increasing the mass inside a black hole by many orders of magnitude.
Accordingly, our own Universe could have been born in the big bounce from the interior of a black hole existing in another universe \cite{infl,BH}.

In a baby universe, the parent universe appears as the other side of the only white hole.
Such a white hole is the boundary of the baby universe.
The baby universe in a black hole is closed (except the white hole connecting it with the parent universe) and can be thought of as a three-dimensional analogue of the two-dimensional surface of a sphere with a hole.
The four-momentum of the gravitational field and matter in such a universe is thus given by (\ref{moment2}), in which the surface integral is taken over the surface of the white hole, and is unchanged by the production of particles near the big bounce.
In the case of a completely closed universe, the surface of its boundary is equal to zero, so the energy and momentum of such a universe are also equal to zero (which has been shown for a homogeneous and isotropic universe in \cite{CI}).

The formation and evolution of a baby universe is not visible for external observers in the parent universe, for whom the event horizon's formation and all subsequent processes would occur after an infinite amount of time had elapsed (because of the time dilation by gravity).
A universe in a black hole is thus a separate, closed region of spacetime with its own timeline.
The motion of matter through the black hole's event horizon can only happen in one direction, providing a past–future asymmetry at the horizon and thus everywhere in the baby universe.
The arrow of time in such a universe would therefore be inherited, through torsion, from the parent universe.
This scenario also solves the black hole information paradox: the information does not disappear in a black hole but goes into the baby universe on the other side of its event horizon \cite{infl}.

The energy $cP_0$ of the gravitational field and matter of a Schwarzschild mass is positive because $m>0$ in (\ref{energy}).
According to the virial theorem, the average kinetic energy $\bar{T}$ of a galaxy is related to its average gravitational potential energy $\bar{U}$ by $2\bar{T}=-\bar{U}$ \cite{LL1}.
If $M$ is the total mass of stars in a galaxy, then $\bar{T}=(1/2)M\bar{v}^2$, where $\bar{v}$ is the root-mean-square stellar speed in the galaxy.
For the observed matter in a galaxy, the ratio of the positive energy from the mass and motion, $Mc^2+\bar{T}$, to the absolute value of the negative gravitational energy, $|\bar{U}|$, is thus equal to $(c^2+\bar{v}^2/2)/\bar{v}^2$.
For the Milky Way, $\bar{v}\approx$ 200 km s$^{-1}$, so such a ratio is on the order of $10^6$ \cite{gal}.
The Milky Way as a whole is moving at a speed of approximately 600 km s$^{-1}$ with respect to the rest frame of the cosmic microwave background.
Such a ratio at extragalactic scales (and in the homogeneous Universe) is thus on the same order.
The positive energy from the mass and motion of the observed matter in the Universe therefore exceeds in magnitude the negative gravitational energy from gravity.
Accordingly, the total energy of the visible matter in the Universe is positive.
However, the total energy in the Universe is given by $cP_0$ with $P^i$ given by (\ref{moment2}) and integrated over the surface of the parent white hole.
The energy of the visible matter therefore exceeds the total energy of the Universe by many orders of magnitude.
Consequently, the Universe must contain another form of matter whose total energy is negative.
This form might account for a large part of the mass that appears to be missing in the Universe \cite{dark} and which is known as dark matter.
In this case, the two problems: the matter compensating for the zero energy of the Universe and dark matter, could be solved together.

If dark matter has a negative energy, it cannot be composed from particles because they have positive rest masses.
It cannot be composed from hypothetical particles with negative rest masses \cite{neg} because they gravitationally repel the ordinary matter and dark matter is gravitationally attractive \cite{dark}.
Analyses of structure on galactic and subgalactic scales have suggested that dark matter may be composed of objects other than weakly interacting subatomic particles \cite{sci}.
To avoid the formation of naked singularities \cite{neg}, dark matter with a negative energy should be coupled to torsion which prevents singularities in fermionic matter \cite{non}.
A possible solution to these conditions could be a scalar field with a negative kinetic term and a torsion-coupling term in the Lagrangian density, decoupled from the electroweak and strong interactions.
Dark matter with a negative energy existing in the closed Universe will be investigated elsewhere.

\end{document}